# Murburn scheme for thermogenesis mediated by uncoupling protein

Kelath Murali Manoj[*1], Daniel Andrew Gideon[2] & Vivian David Jacob[3]

*Satyamjayatu: The Science & Ethics Foundation, Kulappully, Shoranur-2 (PO), Kerala-679122, India*

Corresponding author (KMM): murman@satyamjayatu.com

# Murburn scheme for thermogenesis mediated by uncoupling protein

**Abstract:** Thermogenesis by uncoupling protein (UCP) has traditionally been explained as the dissipation of proton gradient across the inner mitochondrial membrane into heat. However, there are differences of opinion on how thermogenesis is achieved by UCPs and the mechanistic theories have not been correlated sufficiently with UCP's structure. Recent experimental evidence suggests strong correlation of diffusible reactive oxygen species (DROS) with UCP-induced thermogenesis. Further, the mechanistic explanations of mitochondrial oxidative phosphorylation (mOxPhos) were recently revamped with murburn concept, which considers DROS as an obligatory catalytic agent in mOxPhos. Herein, we propose that UCPs (aided by the large pore and positively charged amino acids of suspended loops) enable protonation and transport of DROS. Thus, UCP facilitates DROS-reactions amongst themselves, forming water and liberating heat around the inner mitochondrial membrane. Thereby, the simple murburn scheme for biothermogenesis integrates structural information of UCP with its attributed physiological function.

## Introduction

Brown adipose tissue (BAT) gets its brown color owing to the presence of large amounts of mitochondria, which are in turn thus colored owing to high amounts of heme-containing membrane proteins. In 1978, a ~32 KD protein was implicated in BAT mitochondrial heat generation (thermogenesis). This protein was called uncoupling protein (UCP) (Nicholls & Rial, 1999). Over time, other structural/functional homologs UCPs were discovered (UCP1, UCP2….UCP5), and they were also called mitochondrial anion carrier proteins (MACP). UCP1, also called thermogenin, is now understood to serve as the key player of thermogenesis in BAT. The uncoupling phenomenon mediated by UCPs is envisaged to pose key therapeutic potential and clinical implications in obesity (Chechi et al., 2014), diabetes, and cancer (Rousset et al., 2004). As per the explanations available in textbooks (Berg et al., 2002; Lehninger, Nelson, & Cox, 2004; Voet and Voet, 2011) and reviews (Klingenberg et al., 1999; Kadenbach, 2003; Bouillaud et al., 2016), a proton gradient (or trans-membrane potential, that is otherwise metabolically harnessed to synthesize ATP) gets dissipated into heat generation, aided by the action of UCP. Very recently, some experimental works questioned the classical perception that free fatty acids were essential for UCP-mediated uncoupling (Shin et al., 2017; Schreiber et al., 2017). In recent times, concrete physiological evidences have also accrued for a diffusible reactive oxygen species

(DROS)-based origin for thermogenesis (Chouchani et al., 2016; Kazak et al., 2017; Jastroch, 2017). In parallel, by comparing to the microsomal xenobiotic metabolism (mXM) model (Manoj et al., 2016), we had discredited the long-standing explanations for mitochondrial oxidative phosphorylation (mOxPhos) and proposed a murburn scheme (a newly coined term standing for "**mur**ed/closed **burn**ing", connoting a **m**ild but **u**nrestricted **r**edox catalysis) to explain mOxPhos (Manoj, 2017; Manoj, 2018; Manoj et al., 2018). Keeping the spectrum of such developments in perspective, we perform structure-function analyses of UCPs and propose that thermogenesis sponsored by UCP can be explained by the murburn scheme of mitochondrial metabolism.

**Materials and Methods**

Herein, only direct molecular mechanistic aspects are investigated and indirect/ allosteric regulations, control by nucleotide binding, other metabolic/ hormonal/ neuronal controls and expressional outcomes, etc. are not discussed. Since crystal structures of the native human UCPs are not yet available, a combination of approaches were used to derive structural insights of the trans-membrane helical segments and the coils/loops/strands in matrix/inter-membrane space. In one approach, the amino acid sequences of proteins were retrieved from Uniprot database and aligned using multiple alignment logics (Jalview, CLUSTAL, etc.) and the primary structures were also analyzed for trans-membrane helices/segments (TMH/TMS) identification through online servers (TMpred, TOPCONS, topPRED, PRED-TMR2, etc.). In another approach, the solution-state NMR-determined structure of recombinant murine UCP2 [2LCK (Berardi et al., 2011), of residues 14-309 and a C-terminal $His_6$ tag, with a probe-modified protein bound to GDP, solubilized within DNA nanotubes and phospholipids] was used as a template to derive structural homology models [using Swiss-Model (Bienert et al., 2017) and $Phyre^2$ (Kelley et al, 2015)] of native human UCP1. In a third approach, I-TASSER (Yang et al., 2015) threading strategy was used for modeling human UCP1 structure from a library of proteins' structures and the most suitable model was chosen. A consensus was derived from the three exercises above regarding the tentative location of the six TMS and their delineation from the strands/coils/loops on the matrix/inter-membrane space (IMS). This was also similar to the experimental model proposed by NMR-mapping and homology analyses of human UCP1 structure (Zhao et al., 2017). Therefore, the packing/locations of the TMS and the intervening coils (of matrix and inter-membrane space) were determined with an acceptable level of confidence for human UCP1. The $Phyre^2$ and I-TASSER models of structures of UCPs thus derived were compared to the non-trans-membrane regions of other membrane channels like aquaporin (3GD8, 1JN4) and murine UCP2. The pdb files were viewed and plotted with Pymol/Chimera (DeLano, 2002; Petterson et al., 2004) and cavity analysis was done with CAVER/POCASA (Chovankova et al., 2012; Yu et al., 2010) with probe radius and grid size of 2 Å and 1 Å, respectively.

**Results and Analyses**

Figure 1 and & Table 1 present a brief snapshot of the results obtained in the current study. A comparison of UCP(s) was made with other small membrane proteins, with 5-7 helices. It can be seen that the loops of matrix (and to some extent, the inter-membrane space) of UCP1 showed a preponderance of positively charged residues, lysine (K) and arginine (R). Further, the packing of the six helices in UCP(s) is rather "loose" and the enclosed channel/cavity dimension is far greater than a protein of similar size (with pore), aquaporin. While aquaporin could have a dimension of 3-5 Angstroms as the pore diameter, UCP1's channel was approximately 3 to 5 times greater. The amino acids lining the coils in an anion-transport protein like ADP/ATP porter (1OKC) also showed an excess of lysine and arginine. However, the channel within this protein was smaller and hour-glass shaped (constriction in the middle of the TM region, results not shown).

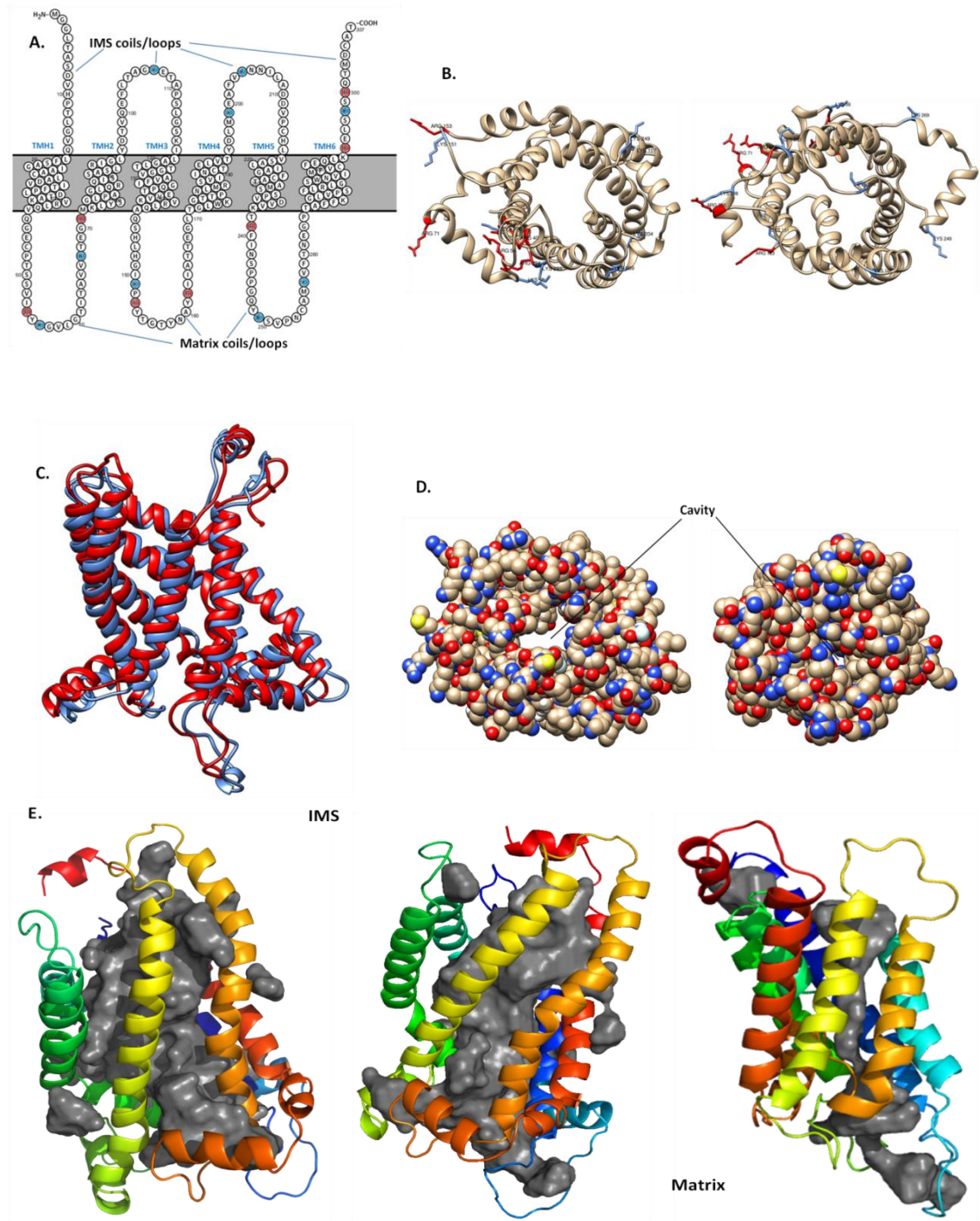

**Figure 1: Images for the structural bioinformatics studies of UCP.** *(a) Protter plot of consensus secondary structure of human UCP1. Lysine and Arginine residues are shaded, and the helical/loop residues demarcated. (b) Comparison of matrix and IMS views of Phyre[2] cartoon structure of UCP1. The Lys and Arg residues are shown with blue and red stick depictions. (c) A juxtaposition of the I-TASSER model of UCP1 (red)*

*with the NMR-derived structure of murine UCP2 (blue) shows that the helical and extra-membrane loops are conserved. (**d**) Matrix views of the channels of Phyre² model of UCP1 and crystal structure of aquaporin. The average pore diameter of UCP1 is clearly much larger than that of aquaporin. (**e**) The lateral views of the Phyre² and I-TASSER models of UCP1 are compared with aquaporin's channel (shown as grey surface) on the right.*

**Table 1: Analysis of loop/coil amino acids and cavity of UCP, with respect to controls.**

| No. | Protein | Pore Volume (Å³)* | Charged amino acids on coils/loops | |
|---|---|---|---|---|
| | *(Amino acids, TM helices)* | POCASA, CAVER | *Matrix (+ve, -ve)* | *IMS (+ve, -ve)* |
| 1 | **Murine mitochondrial translocator (2MGY) (169, 5)** | No pore | K39, R103, R161, R162, R166, E169 (5, 1) | E3, R128 (1, 1) |
| 2 | **Human A2A receptor (3PWH) (329, 7)** | 392 (conical), na | E212 (0, 1) | D261 (0, 1) |
| 3 | **Bovine aquaporin (1J4N) (271, 6+2)** | 439, 509 | K40, D50, D123, D210 (1, 3) | E4, D160, R161, R162, R163, R164, D165, R236, D239, D242, R243, K245 (7, 5) |
| 4 | **Human aquaporin (3GD8) (223, 6+2)** | 402, 366 | R108, K109, D179, D181, D182, D184 (2, 4) | E63, K64, D69, D228 (1, 3) |
| 5 | **Human UCP1 (SwissModel) (307, 6)** | 5209, na | R54, K56, R71, K151, R153, K249, K257, K269 (8, 0) | K107, K204, D210, D211, K293, R294, E295, K298, R300, D304 (6, 4) |
| 6 | **Human UCP1 (Phyre²) (307, 6)** | 5534, na | R40, R54, K56, K151, R153, K249, K269 (7, 0) | K107, K204, D210, D211, K293, R294, E295, K298, R300, D304 (6, 4) |
| 7 | **Human UCP1 (I-TASSER) (307, 6)** | 4050, na | R40, R54, K56, R71, K151, R153, K249, K257, K269 (9, 0) | D8, K107, D210, D211, K204, K293, R294, E295, R300, D304 (5, 5) |
| 8 | **Murine UCP2 (2LCK) (303, 6)** | 4930, na | E46, R52, R60, R149, R154, R155, R262, K263, E264 (7, 2) | K109, E112, K206, D212, D213, R305, E306 (3, 4) |

*\* Assuming a grossly cylindrical pore for a phospholipid membrane 30 nm thick, a mean diameter of 4-5 Å gives a hydrodynamic pore volume of 375-590 Å³ for aquaporin. Assuming a mean diameter of 13-15 Å, UCP1 would have a pore volume of 4000-5300 Å³. This means that UCP's channel volume is approximately 10 times aquaporin's, implying a much higher flux density across the UCP.*

**Discussion**

Even though there are several mechanistic proposals for thermogenesis by UCP1, they can be broadly addressed (Figure 2) under three models/types. In the fatty acid protonophore model (Skulachev, 1991; Garlid et al, 1996), UCP does not deal with protons at all and only helps the free fatty acyl anion flip directionally from the matrix side to the inter-membrane side. Then, the protonated fatty acid flips from the inter-membrane side to the matrix-side on its own. In the proton buffering model (Winkler & Klinkenberg, 1994; Rial et al., 2004), UCP1 acts as a proton carrier (directly or indirectly), assisted by the fatty acid. In these proposals, a UCP-bound fatty acyl group helps in swinging the proton from the IMS (inter-membrane space) to the matrix side. In the symporter model (Fedorenko et al., 2012), the proton is transported along with the fatty acid through the pore of UCP. Essentially, all the three UCP-sponsored thermogenesis classical mechanisms obligatorily require fatty acids and thermogenesis results owing to the movement of inter-membrane space (IMS) protons (that were originally expelled by the purported ETC-proton pumps). That is, a matrix-ward movement of IMS protons is facilitated by the UCP and this process undoes the built-up proton-gradient (or trans-membrane potential), and in turn, this "potential dissipation" manifests as heat (Figure 2, step 3).

All the three classical mechanistic models for UPC-mediated thermogenesis counter key aspects of mitochondrial physiology and thermodynamics. The oxidation of NADH/succinate mediated by the inner-membrane proteins of mitochondria would not involve matrix to IMS proton-pumping because mitochondria lack the required amount of protons (Manoj, 2017; Manoj, 2018). Therefore, a proton-pump based potential gradient cannot be built from a starting state and cannot operate in a steady state. Even if proton pumps were to be involved in mitochondria, the outward pumping process would be exothermic (as oxidation of NADH is exothermic), which must directly imply that the inward movement across the same membrane-system must be endothermic. So, a purported inward movement of the pumped-out inter-membrane space protons, generating heat, is not thermodynamically feasible. Chemically, the outcomes (disconnection of ATP-synthesis from NADH-oxidation, with the concomitant generation of heat) mediated by UCP are quite similar to what results from the purported flip-flop motions that dintrophenol (DNP) makes (as an uncoupler) in mitochondrial membrane. If mitochondrial membrane was impervious to protons (a small unit positive charge), it is improbable that the membrane would be freely permeable to dinitrophenolate ion, which is a much larger molecule with three negative and two positive charges. The surface of inner mitochondrial membrane is highly dense with negative charges, owing to the preponderance of cardiolipin. So, a spontaneous or directional flip-flop movement of nitrophenolate anion (and in analogy, fatty acyl anion) across the membrane is infeasible. Yet, DNP is known to uncouple NADH-oxidation from ATP-synthesis, leading to thermogenesis. Since there is little within the chemical structures of DNP/UCP that is capable of generating heat by its own merit, the mechanism of thermogenesis would not depend on sophisticated structural aspect of the membrane or trans-membrane potential. Therefore, it is reasonable to deduce that the cause for thermogenesis must primarily lie in mitochondrial metabolism and how DNP/UCP affects the dynamics of the metabolic process.

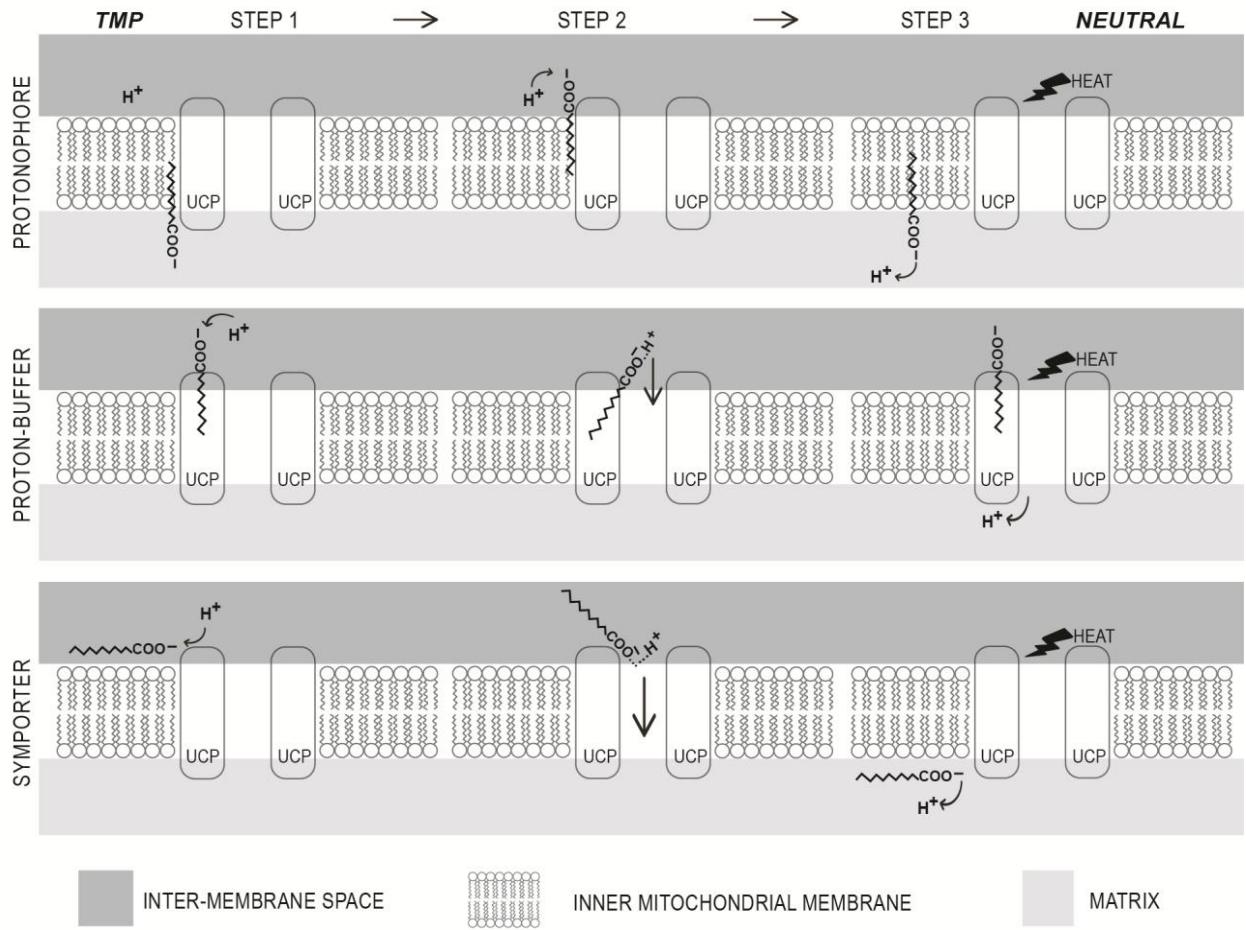

*Figure 2: The various classical models of UCP-mediated thermogenesis in mitochondria:* [For details of mechanism, please refer text.]

Recently, the classical mechanism(s) has been countered by the observation that UCP-mediated uncoupling was observed in BAT even without lipolysis (Shin et al., 2017; Schreiber et al., 2017). So, it is imperative to think beyond the classical paradigms of uncoupling mediated by UCP. In the last five years, experimental research has opened up new perspectives towards the roles of ROS in mitochondrial metabolism. Recent experimental works with the BAT *in vivo* systems have provided substantial evidence that thermogenesis occurs through DROS dynamics (Chouchani et al., 2016; Kazak et al., 2017; Jastroch, 2017). DROS is a key causative agent that leads to heat production in brown adipose tissue in methamphetamine induced hyperthermia models (Sanchez-Alvarez et al., 2014). In experimental studies, it was shown that UCP knocked-out mice suffered from elevated levels of oxidative stress in cold (Stier et al., 2014). These studies demonstrate that DROS dynamics is obligatorily connected to physiological thermogenesis. [Earlier experimental evidence also indicates the interaction of ROS with fatty acids and UCP (Echtay et al., 2002). UCPs 2 & 3 are supposedly involved in ROS control and mild uncoupling is supposed to prevent ROS production (Criscuolo, 2006). Evidence for activation of proton conductance by ROS-associated roles of UCP was also documented (Talbot et al., 2003).]

The ability of charged amino-acids to modulate DROS (conversion of superoxide to peroxide or vice versa) has been clearly established in redox-active membrane proteins.

It was found in an NADPH oxidase enzyme that deletion of a loop or even the mutation of a single histidine residue or synthetic peptides with positive amino acids could alter the DROS-modulating ability of membrane-bound protein (El-Banna et al., 2010; Joseph et al., 1994; Takac et al., 2011). An analogy from the mXM system is very useful to understand the uncoupling by UCP in mOxPhos. It was observed in the mXM system that when the N-term segment (rich in positively-charged residues) of flavoenzyme reductase was truncated and left within the reaction milieu, high levels of uncoupling/inhibition of cytochrome P450 (CYP) mediated reactions were noted. We had reasoned that this resulted because the N-term segment is capable of DROS modulation, leading to lowered lifetime of the catalytically useful superoxide. Otherwise, the intact N–term segment served as an anchor for the protein for the phospholipid membrane, a feature which enabled enhanced catalysis by the CYP because membrane anchoring and proximity to reductase would enable efficient harnessing of DROS (Gideon et al., 2012). Analogously, as shown in this study, the coils (in matrix and IMS) of UCP harbor lysine-arginine rich residues and coupled with the large trans-membrane channel, these proteins could affect the bulk DROS-dynamics within the mitochondrial system. There is no direct chemico-physical evidence or structural correlation for how trans-membrane proton movement (which is expected to occur everywhere in cells!) could lead to ATP-synthesis or heat generation. However, the simple chemical reaction involving DROS are well-known to generate heat and DROS-dynamics can also be correlated to UCP structure. We have already demonstrated that superoxide could aid ATP synthesis (Manoj et al., 2018). Therefore, the murburn model of uncoupling, as shown in Figure 3 could help explain the phenomenon of heat generation in BAT. The wide pore of UCP could effectively serve as (aquated) proton and/or DROS high flux-density conduit and the positively charged amino acids lining the UCP channel could help protonate a DROS like superoxide. Recent studies have revealed that UCPs can serve as generic anion transporters (Monne et al., 2018). In turn, this would facilitate more radicals reacting among themselves (leading to water formation) on both sides of the inner membrane, consequently generating heat. (Since the IMS is cramped and proton availability is high, heat generation would be efficient.) Fatty acids are interface-active agents that would be expected to affect the dynamics of DROS in a concentration-dependent manner, which could explain the differences that prevailed in the field. Since the transient concentration of the negatively charged superoxide goes down in the matrix owing to the uncoupling process, the trans-membrane potential is also diminished as a consequence. An analogy is again relevant here from the mXM system- interfacial DROS-scavengers/modulatos (dihalophenolics and amphipathic vitamins/derives, respectively akin to DNP and fatty acid) were misunderstood to be active-site inhibitors of CYP enzymes (analogous to the "proton-potential" of mOxPhos). We had shown that such molecules affected the DROS dynamics at the phospholipid interface, where CYP and the flavoenzyme NADPH-oxidase were co-localized (Manoj et al., 2010; Gideon et al., 2012, Manoj, Gade, et al, 2016; Manoj, Parashar, et al., 2016; Parashar et al., 2014; Parashar et al., 2018). Therefore, the effects of DNP-mediated uncoupling and thermogenesis in mOxPhos is explained by the DROS-modulation ability (not proton-potential dissipation!) of membrane bound DNP.

The critical dependence of rates and energetics on protons can be well understood by considering the disproportionation and one-electron redox equilibrium of a key element of murburn scheme- superoxide. Some reactions are shown in Box 1 below; to highlight the concepts of chemistry involved (Sawyer & Valentine, 1981, Kanematsu & Asada 1994).

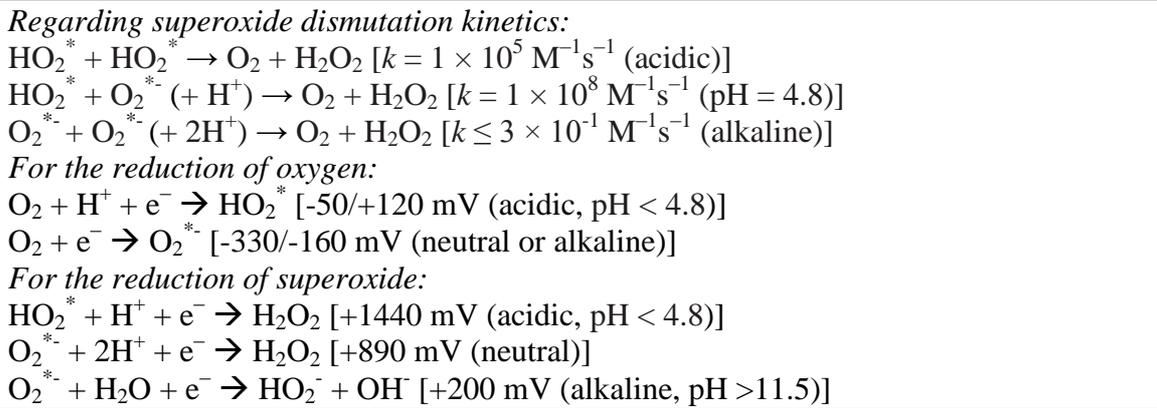

*Regarding superoxide dismutation kinetics:*
$HO_2^* + HO_2^* \rightarrow O_2 + H_2O_2$ [$k = 1 \times 10^5$ M$^{-1}$s$^{-1}$ (acidic)]
$HO_2^* + O_2^{*-} (+ H^+) \rightarrow O_2 + H_2O_2$ [$k = 1 \times 10^8$ M$^{-1}$s$^{-1}$ (pH = 4.8)]
$O_2^{*-} + O_2^{*-} (+ 2H^+) \rightarrow O_2 + H_2O_2$ [$k \leq 3 \times 10^{-1}$ M$^{-1}$s$^{-1}$ (alkaline)]
*For the reduction of oxygen:*
$O_2 + H^+ + e^- \rightarrow HO_2^*$ [-50/+120 mV (acidic, pH < 4.8)]
$O_2 + e^- \rightarrow O_2^{*-}$ [-330/-160 mV (neutral or alkaline)]
*For the reduction of superoxide:*
$HO_2^* + H^+ + e^- \rightarrow H_2O_2$ [+1440 mV (acidic, pH < 4.8)]
$O_2^{*-} + 2H^+ + e^- \rightarrow H_2O_2$ [+890 mV (neutral)]
$O_2^{*-} + H_2O + e^- \rightarrow HO_2^- + OH^-$ [+200 mV (alkaline, pH >11.5)]

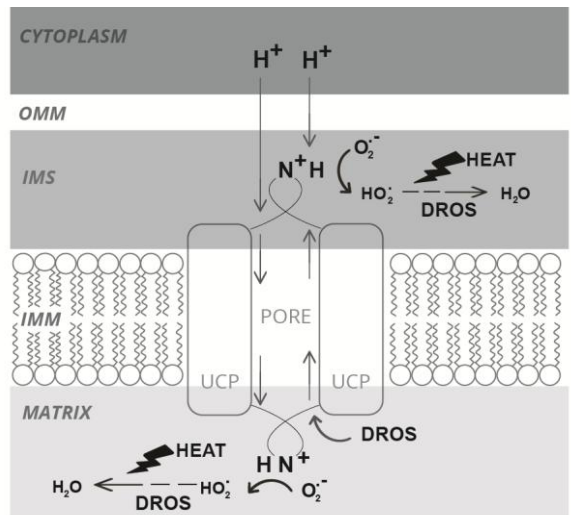

*Figure 3: The murburn scheme for UCP-aided thermogenesis in mitochondria: [Key: OMM- outer mitochondrial membrane, IMS- inter-membrane space, IMM- inner mitochondrial membrane] There are two modalities of uncoupling and heat generation- one is by matrix-ward inundation of protons and the other is via IMS-ward movement of DROS.*

As seen, protons play a critical role in several steps and affect both the thermodynamic and kinetic aspects (and this also explains the historical confusions regarding proton involvement). Superoxide is the key agent for the initiation of the oxygen-centric murburn mOxPhos scheme. Although superoxide production would be independent of protons, superoxide activity (for ATP-synthesis) would be critically dependent on protons because its protonated species (hydroperoxyl radical) is known to be many times more (re)active than superoxide itself. If the superoxide gets protonated near the membrane, it would be easily lost to Haber-Weiss chemistry in the aqueous milieu (leading to the liberation of heat energy). Therefore, murburn scheme explains the wide-bore channel and the extensive distribution of charged amino acids on the loops of UPC, both of which were unaccounted by earlier mechanistic proposals. The double membrane structure of mitochondria can now be seen a feature that enables a "mild and unrestricted" burning within the vicinity of the organelle's inner membrane. Since it is known that NADH oxidation is channelized into ATP synthesized + heat generated, the findings herein serve as support for the obligatory involvement of DROS in the mOxPhos scheme (Manoj et al., 2018).

**Acknowledgments**: This work was powered by Satyamjayatu: The Science and Ethics Foundation. KMM dedicates this manuscript to the fond memories of late Lowell P. Hager (UIUC, USA).

**Conflict of interests**: There are no conflicts of interests to disclose.